\documentclass[english,twocolumn, superscriptaddress,prl]{revtex4-1}
\usepackage[T1]{fontenc}
\usepackage[latin9]{inputenc}
\usepackage{amsmath}
\usepackage{amssymb}
\usepackage{graphicx}
\usepackage{nicefrac}

\makeatletter

\newcommand{\ket}[1]{\left| {#1} \right\rangle}
\newcommand{\bra}[1]{\left\langle {#1}\right|}

\renewcommand{\t}[1]{\textrm{#1}}

\makeatother

\usepackage{babel}
\begin{document}

\title{Beating the Rayleigh limit using two-photon interference}

\author{Micha\l\ Parniak}
\email{michal.parniak@fuw.edu.pl}
\affiliation{Faculty of Physics, University of Warsaw, Pasteura 5, 02-093 Warsaw, Poland}
\affiliation{Centre for Quantum Optical Technologies, Centre of New Technologies, University of Warsaw, Banacha 2c, 02-097 Warsaw, Poland}
\author{Sebastian Bor\'{o}wka}
\affiliation{Faculty of Physics, University of Warsaw, Pasteura 5, 02-093 Warsaw, Poland}
\author{Kajetan Boroszko}
\affiliation{Faculty of Physics, University of Warsaw, Pasteura 5, 02-093 Warsaw, Poland}
\author{Wojciech Wasilewski}
\affiliation{Faculty of Physics, University of Warsaw, Pasteura 5, 02-093 Warsaw, Poland}
\affiliation{Centre for Quantum Optical Technologies, Centre of New Technologies, University of Warsaw, Banacha 2c, 02-097 Warsaw, Poland}
\author{Konrad Banaszek}
\affiliation{Faculty of Physics, University of Warsaw, Pasteura 5, 02-093 Warsaw, Poland}
\affiliation{Centre for Quantum Optical Technologies, Centre of New Technologies, University of Warsaw, Banacha 2c, 02-097 Warsaw, Poland}
\author{Rafa\l\ Demkowicz-Dobrza\'{n}ski}
\affiliation{Faculty of Physics, University of Warsaw, Pasteura 5, 02-093 Warsaw, Poland}
\begin{abstract}
Multiparameter estimation theory offers a general framework to explore imaging techniques beyond the Rayleigh limit. While optimal measurements of single parameters characterizing a composite light source are now well understood, simultaneous determination of multiple parameters poses a much greater challenge that in general requires implementation of collective measurements. Here we show, theoretically and experimentally, that Hong-Ou-Mandel interference followed by spatially resolved detection of individual photons provides precise information on both the separation and the centroid for a pair of point emitters, avoiding trade-offs inherent to single-photon measurements.
\end{abstract}
\maketitle
Multiparameter quantum estimation emerges as a general framework to optimize information retrieval in a variety of experimental scenarios. The problem of imaging
can be viewed as an important example of such a scenario, where the properties of an image, for example locations and intensities of point emitters or
the moments of the image intensity distribution are the parameters to be estimated \cite{Rehacek2017,Paur2016, Yang2016,Chrostowski2017,Ragy2016,Napoli2018,Zhou2018}. A recently introduced family of superresolution imaging schemes \cite{Tsang2009,Tsang2015,Tsang2016,Tsang2017,Kerviche2017,Dutton2018} based on spatial demultiplexing enable one to determine the separation of two nearby point sources below the Rayleigh limit, but
require in principle perfect knowledge of the centroid \cite{Chrostowski2017}. Moreover, at the single photon level they are fundamentally incompatible with the measurement needed to estimate the centroid itself. Nonetheless, the effort to extract optimally information carried in light emitted naturally by a source \cite{Puschmann2005,Tang2016,Nair2016,Nair2015,Lupo2016,Dutton2018,Tham2017} may open up new applications compared to established approaches that require manipulations of the sample to be imaged \cite{Moerner2015}.

A deeper insight rooted in the multiparameter estimation theory reveals a possible solution of the above incompatibility problem. Interestingly, in the strong subdiffraction regime where images of the sources overlap significantly, the problem can modelled as  simultaneous estimation of the
 length and the rotation angle of a qubit Bloch vector \cite{Chrostowski2017}. From the theory of multiparameter estimation it then follows that, provided collective measurement on the photons (or qubits) are allowed, the incompatibility between the optimal individual measurements to estimate the centroid and the sources separation ceases to be an issue \cite{Bagan2006, Vidrighin2014}. The question is how to realize such a collective measurement in practice.

In this Letter we exploit the advantages offered by the multiphoton interference approach, demonstrating
a two-photon protocol for imaging of two point sources, where the centroid estimation is performed in the optimal way, and at the same time the sources separation parameter is estimated with a superresolution precision.
The idea relies on the effect
of two-photon interference and does not require pre-estimation of the centroid or fine-tuning
of the measurement basis inherent to spatial mode demultiplexing schemes \cite{Tsang2009,Tsang2015,Tsang2016,Tsang2017,Kerviche2017,Dutton2018}, where
any systematic error in centroid estimation propagates to separation estimation and significantly degrades the imaging protocol.

In Fig. 1(a) we depict a scenario where two photons emitted by a composite source arrive simultaneously at the input ports of the beamsplitter. The proposed protocol exploits both
cross-coincidences between the output ports and double events in each port, detected with spatial resolution \cite{Jachura2016}. The number of cross-coincidences grows with the distinguishability of the two photons and therefore carries information about the separation between point sources. 
Most importantly, the proposed
interferometric scheme does not require prior selection of the measurement basis or
the axis of symmetry, as the two photons serve as a reference for each
other. Furthermore, thanks to spatially-resolved detection this strategy will be shown to be robust against residual spectral distinguishability. Let us note that previous approaches to collective
measurements relied on the fundamental advantage of using photonic entanglement \cite{Rozema2014}, also for superresolution photolithography \cite{Liao2010,Boyd2012,DAngelo2001,Boto2000},
which is essentially different from our technique of simply utilizing the
bosonic nature of photons.

The somewhat non-trivial demand of interfering two photons from a realistic classical (thermal) composite source on a beamsplitter could be realized by a photon number quantum nondemolition (QND)
measuring device that preserve
spatial properties of light, and upon registering single photons delays and redirects them so that they arrive together at the two beamsplitter input ports. Recent advances in storing and controlling single
photons in quantum nonlinear media such as Rydberg atoms \cite{Firstenberg2013} as well as spatially-multimode quantum memories \cite{Parniak2017} with processing capabilities \cite{Parniak2018} could provide
a viable way to realize the scheme. In particular, a $\pi$ phase shift induced by a single photon has already been achieved \cite{Tiarks2016} and current experiments already explore the Rydberg interactions in the trasverse spatial domain \cite{Busche2017}. The combination of a multimode quantum memory with the spatially-resolving QND measurement could follow the steps of experiments demonstrating optical storage in Rydberg media \cite{Distante2017,Li2016}, use alternative proposals such as nonlinearities induced by ac-Stark shifts \cite{Everett2016} or utilize novel solid-state systems with similar capabilities yet broader spectral bandwidths \cite{Yang2018a,Walther2018}.
\begin{figure}
\includegraphics[width=\columnwidth]{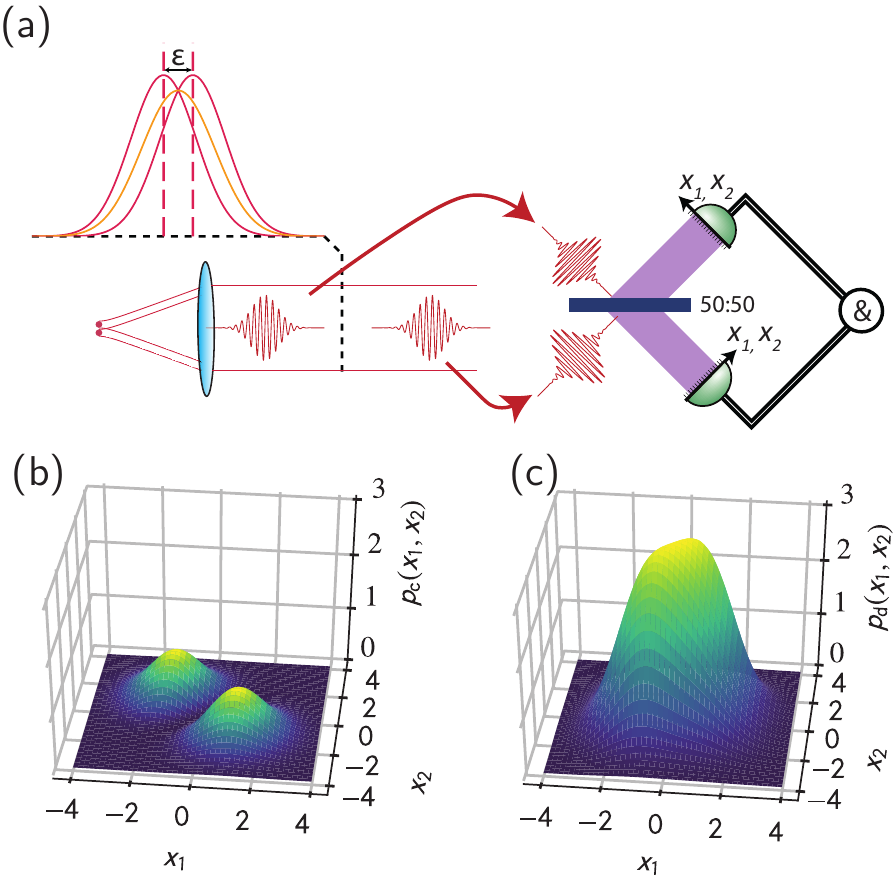}
\caption{The idea for collectively-enhanced quantum imaging protocol. (a) Two point source are imaged using an optical system
with a well-defined intensity point-spread function (inset, all curves are normalized to equal integrals). The photons are made to interfere (see text for details) at the output beamsplitter, after which we register cross-coincidences and double events with spatial resolution. Information about sources separation $\varepsilon$ as well as centroid $x_0$ are drawn both from the ratio of cross-coincidences (b) to double-events (c) and their spatial probability distributions $p_c(x_1,x_2)$ and $p_d(x_1,x_2)$ (here expressed in arbitrary units).
}
\end{figure}

To support the intuitions behind the discussed scheme let us compare
the two-photon imaging scheme with direct imaging (DI) by modeling
a problem of resolving a 1D image formed by two point sources.
Let $\psi(x-x_0)$ be a 1D wave function representing the amplitude transfer
function of a single source in the image plane centered at point $x_0$.
We assume
that this function is determined by well characterized properties of the imaging
setup.
In what follows we denote the corresponding single photon state characterized by $\psi(x-x_0)$ as $\ket{x_0}$.

Consider a situation where the image is produced as
a result of an incoherent overlap of images of two point sources separated
by a distance $\varepsilon$, located at $x_{+}=x_0+\varepsilon/2$
and $x_{-}=x_0-\varepsilon/2$.
We may then write the spatial density matrix of a photon
emitted from the system as $\rho=1/2(\ket{x_{+}}\bra{x_{+}}+\ket{x_{-}}\bra{x_{-}})$.

In the DI scheme the probability distribution for the position of the detected photon is given by $p_{\boldsymbol{\theta}}(x)=\tfrac{1}{2}|\psi(x-x_{+})|^{2}+\frac{{1}}{2}|\psi(x-x_{-})|^{2}$,
where $\boldsymbol{\theta}=((x_++x_-)/2,x_+-x_-)=(x_0,\varepsilon)$ represents the dependence on the estimated parameters.
For any locally unbiased estimator, the covariance matrix for the estimated parameters can be lower bounded using the
Cram{\'e}r-Rao inequality \cite{Kay1993}:
\begin{equation}
\mathrm{Cov}\boldsymbol{\theta}\geq \frac{F^{-1}}{N},
\ F_{ij}=\int_{-\infty}^{\infty}\t{d}x\frac{\partial_{\theta_{i}}p_{\boldsymbol{\theta}}(x)\partial_{\theta_{j}}p_{\boldsymbol{\theta}}(x)}{p_{\boldsymbol{\theta}}(x)} \label{eq:fim},
\end{equation}
where $F_{ij}$ is the Fisher information (FI) matrix per single photon, while $N$ represents the total number of photons registered.
The bound is asymptotically saturable using e.g. max-likelihood estimator, hence $\lim_{N \rightarrow \infty} N \mathrm{Cov}\boldsymbol{\theta} = F^{-1}$. As the FI matrix is diagonal for the given problem, we can easily calculate the variances $\Delta^2x_{0} = (F^{-1})_{11}$,
$\Delta^2\varepsilon = (F^{-1})_{22}$ the respective variances of the estimated parameters per single photon used.
In case of DI the FI matrix yields the following precision for estimation in the leading order in $\varepsilon$:
\begin{align}
(\Delta^2x_0)^{-1}_\mathrm{DI} = 1-\frac{\varepsilon^2}{4}, \\
(\Delta^2\varepsilon)^{-1}_\mathrm{DI} = \frac{\varepsilon^2}{8},
\end{align}
where for concreteness we have assumed a Gaussian-shaped transfer function
$\psi(x)=(2\pi)^{-1/4}\exp(-x^{2}/4)$, yielding intensity profile with standard deviation $1$ which can be regarded as a natural unit of distance in the problem.
The above expansion is valid for small $\varepsilon$ when source point images
are separated by a distance smaller than the transfer function spread, and clearly shows impossibility of precise estimation of $\varepsilon$ in the $\varepsilon \rightarrow 0$ limit.

Crucially, as observed in \cite{Tsang2016}, a more fundamental bound based on the quantum FI matrix $F^Q$ \cite{Helstrom1976}, which
does not assume any particular measurement strategy and is based solely on the properties of the quantum state $\rho$ to be measured reads:
\begin{align}
\label{eq:qfi}
(\Delta^2x_0)^{-1}_\mathrm{Q} = 1-\frac{\varepsilon^2}{4}, \\
(\Delta^2\varepsilon)^{-1}_\mathrm{Q} = \frac{1}{4},
\end{align}
indicating a potential spectacular robustness of $\varepsilon$ estimation as the $\Delta^2 \varepsilon$ is constant irrespectively of how small $\varepsilon$ is. While the bound \eqref{eq:fim} with $F$ being replaced by $F^Q$ is
saturable for the problem considered, it requires collective measurements on many copies of $\rho$ \cite{Vidrighin2014,  Rehacek2017, Chrostowski2017, Ragy2016}.

\begin{figure*}
\includegraphics[width=\textwidth]{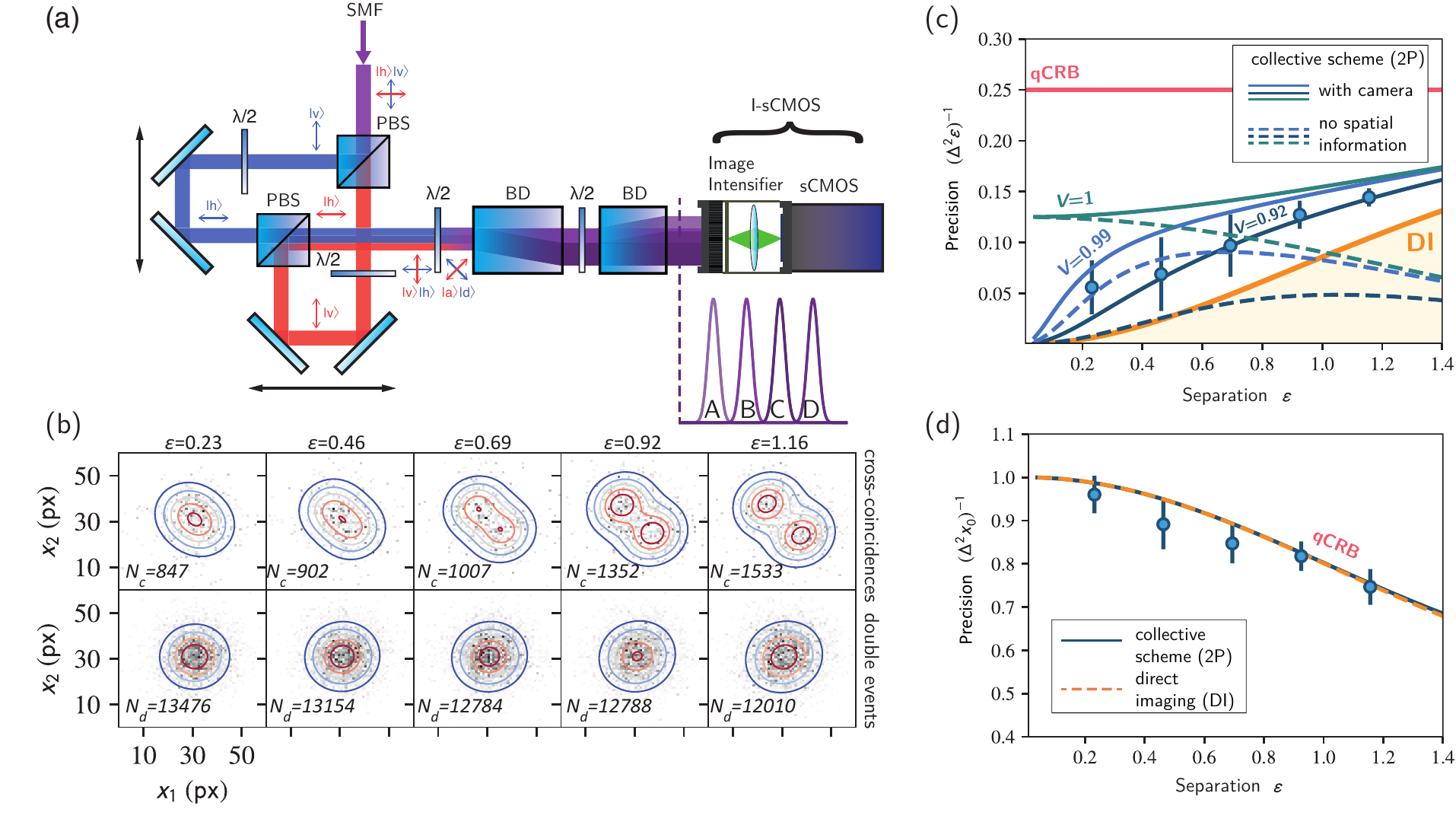}

\caption{Results of the multi-parameter quantum estimation in a super-resolution imaging scenario with a photon-pair
source. (a) Experimental setup for generating a pair of photons in
two adjacent modes (PBS, polarizing beamsplitter; $\lambda/2$, half-wave plate; BD, calcite beam displacer). Using an $|hv\rangle$ photon pair and reconfiguring the positions of the retro-reflectors in the interferometer we generate
the two-photon state expected in the imaging experiment for a set
of values of source separation $\varepsilon$. The output single mode
fiber (SMF) face is imaged onto the I-sCMOS sensor (see Supplementary Material or Ref. \cite{Chrapkiewicz2016} for details of I-sCMOS sensor operation and construction) photocathode so
that the beam has a flat wavefront with $1/e^2$ diameter of $100\ \mu\mathrm{m}$. The camera registers cross-coincidences
(as coincidences between regions A-C, A-D, B-C and B-D) and double
events (as coincidences between regions A-B or C-D). (b) Spatially-resolved cross-coincidences
(top) and double events (bottom) along with fitted model with Gaussian mode shape for subsequent values of $\varepsilon$ corresponding to data point in panels (c) and (d). Color scale for each map is normalized separately to highlight shape intricacies. (c) Precision of estimation of $\varepsilon$ for the $\rho^{\otimes2}$ state for and (d) precision of estimation of the centroid position $x_0$ as a function of source separation $\varepsilon$. The collective 2P scheme provides an enhancement in estimation of $\varepsilon$ while preserving the precision of centroid estimation. The ultimate precision limit given by the quantum Cram{\'e}r-Rao bound is denoted by qCRB, and for the precision of centroid estimation in (d) it overlaps with the precision obtained with the protocols we employ. Theoretical curves are obtained from numerically evaluated FI. Errorbars correspond to one standard deviation of results obtained from each dataset containing 1000 coincidences (see Supplementary Material for details).}
\end{figure*}

We are now ready to quantify the precision of
estimating $x_0$ and $\varepsilon$ in the two-photon (2P) interferometric scheme and contrast it with the above-mentioned strategies.
Given $\rho^{\otimes 2}$ at the input ports of the beam-splitter, we calculate
spatially-resolved propabilities for coincidences $p_{\mathrm{c}}(x_{1},x_{2})$ as well as double events
$p_{\mathrm{{d}}}(x_{1},x_{2})$, from which the information about $x_0$ and $\varepsilon$ is drawn. Furthermore, we assume a known two-photon visibility $\mathcal{V}$ resulting from the operation of the non-demolition photon routing device before the beamsplitter.
The resulting precision of estimation per single photon used, see Supplementary Material, expanded up to the second order in $\varepsilon$ reads:
\begin{align}
\label{eq:2p-1}
(\Delta^2x_0)^{-1}_\mathrm{2P} &= 1-\frac{\varepsilon^2}{4}, \\
(\Delta^2\varepsilon)^{-1}_\mathrm{2P} &= \begin{cases} \frac{1}{8} + \frac{5}{128} \varepsilon^2 &
\label{eq:2p-2}
\mathcal{V} = 1  \\ \frac{4-\mathcal{V}^2}{32(1-\mathcal{V}^2)} \varepsilon^2  & \mathcal{V} < 1 \end{cases},
\end{align}
while the expansion in case of imperfect visibility is valid in the regime where $\varepsilon^{2}\lesssim1-\mathcal{{V}}$.
In case of perfect interference, we see that while keeping the optimality of $x_0$ estimation, we additionally obtain $\varepsilon$ estimation with precision reduced by approximately a factor of $2$ compared to the fundamental bound given in (5).
This shows superiority of 2P over DI, with the added advantage that the measurement setting is fixed and
does not require adjusting the measurement for $\varepsilon$ depending on preestimation of $x_0$.
Here we would like to stress the importance of spatial information that is available in the experiment: if only the ratio of coincidence and double events was available, there would be no information on $x_0$ parameter at all, while the precision of $\varepsilon$ estimation
shows a small reduction for finite $\varepsilon$ compared with Eq. \ref{eq:2p-2}  and reads: $\frac{1}{8} - \frac{5}{128} \varepsilon^2 + O(\varepsilon^4)$ when the visibility is equal to one.

The role of spatial information becomes more pronounced for finite visibilities $\mathcal{V}$, for which the spatial information always provides an advantage for all values of $\varepsilon$ compared to the case when we consider only the ratio of cross-coincidences and double-event where the precision reads $\mathcal{V}^2  [32(1-\mathcal{V}^2)]^{-1} \varepsilon^2+ O(\varepsilon^4)$. This is achieved as coincidences that arise due to finite visibility are characterized by a different spatial distribution than coincidences that are due to spatial separation. In both cases we recover the $\varepsilon^{2}$-scaling and thus for
small $\varepsilon$ the advantage of the collective schemes over
DI takes the form of a constant factor rather than favorable scaling. Nonetheless, as
this factor scales as $(1-\mathcal{V})^{-1}$ the enhancement can
be significant.

For a proof-of-principle experimental demonstration, we generated families of states $\rho^{\otimes 2}$ for a set of separations
$\varepsilon$ (see Fig. 2(a) and Supplementary Material for details of the interferometric setup). In Fig. 2(c) and 2(d) we plot the final precision of estimation
divided by the total number of photons used as a function of $\varepsilon$ (see Supplementary Material for details of data analysis). The proposed theory (for $\mathcal{{V}}=0.92$) accurately predicts the estimation precision
for the given experimental parameters demonstrating a significant, over twofold
enhancement over the DI scheme. The spatial resolution provides an advantage over the whole range of parameters, as it allows us to distinguish effects of finite visibility versus the reduced mode overlap due to source separation.

In Figure 2(c),(d) we additionally plot the theoretical predictions for $\mathcal{{V}}=0.99$ and perfect interference
i.e. $\mathcal{{V}}=1$.
The precision approaches a constant value for $\varepsilon\rightarrow0$
only for $\mathcal{V}=1$, but offers significant enhancement for
realistic visibilities. Note that if information is drawn only from the number of coincidences
to double events with no spatial resolution, we can still beat the DI scheme
over a broad range of parameters, especially for small $\varepsilon$.
This highlights the possibility to perform precise imaging with only
single-pixel detectors.

Let us now provide a simple argument for the observed degree of precision enhancement. 
The approximately two-fold reduction of precision for $\varepsilon$ estimation for $\mathcal{V}=1$ in the 2P protocol compared to the fundamental bound is due to the fact that the protocol performs collective measurement on two photons only. The essence of the collective measurement is effective projection of $\rho^{\otimes 2}$ on symmetric and antisymmetric subspaces thanks to the properties of the Hong-Ou-Mandel interference. Such a measurement commutes with
joint unitary rotation of the state $U^{\otimes 2} \rho^{\otimes 2} U^{\dagger \otimes 2}$  which represents the shift of the centroid $x_0$ in our model, and hence
does not collide with the  measurement optimal for extracting information on $x_0$.
Theoretically, if collective measurements on arbitrary number of copies were possible, one could
project $\rho^{\otimes N}$ state on subspaces corresponding to different irreducible representation  of the permutation group
 which provides optimal information about the $\varepsilon$ parameter in the $N \rightarrow \infty$ limit and does not interfere with the optimal measurement of $x_0$ \cite{Bagan2006}. Thus, through harnessing more than two photons one would be able to approach and even saturate the quantum Cram{\'e}r-Rao bound \eqref{eq:qfi}.

 Interestingly, in a slightly modified imaging scenario, the two-photon measurements may actually saturate the limit discussed above. Consider a different variant of the two-photon state impinging on the beamsplitter
   \begin{equation}
   \rho_{11}=\frac{{1}}{2}(\ket{x_{+}}\bra{x_{+}}\otimes\ket{x_{-}}\bra{x_{-}}+\ket{x_{-}}\bra{x_{-}}\otimes\ket{x_{+}}\bra{x_{+}}),
   \end{equation}
which represents a situation where the photons from the two sources always enter at different input ports of the beamsplitter.
Such a two-photon state could be obtained from a pair
of single-photon emitters excited simultaneously, where we would never observe two photons emitted from the same source.
In this case,
 analogous calculations to the ones presented in Ref. \cite{Rehacek2017} for the $\rho$ state, lead to the quantum FI matrix
  which in the leading order in $\varepsilon$ remains unchanged, whereas the two photon experiment described above saturates the bound exactly:
  \begin{align}
\label{eq:2p-3}
(\Delta^2x_0)^{-1}_{\mathrm{2P},\rho_{11}} = (\Delta^2x_0)^{-1}_{\mathrm{Q},\rho_{11}} = 1, \\
\label{eq:2p-4}
(\Delta^2\varepsilon)^{-1}_{\mathrm{2P},\rho_{11}}= (\Delta^2\varepsilon)^{-1}_{\mathrm{Q},\rho_{11}} =\frac{1}{4}.
\end{align}

Finally, it is insightful to juxtapose the presented scheme with the 
celebrated Hanbury Brown--Twiss (HBT) interferometry
\cite{Hanbury1954, Hanbury1956, Brown1957, Brown1958, Fano1961}. The 
essential difference is that in our approach photon positions are 
measured in the image plane, while in the HBT scenario spatially 
resolved detection is implemented in the Fourier plane conjugate to the 
source. For photons arriving from point sources located at angular 
positions specified by wave vectors ${\bf k}_1$ and ${\bf k}_2$ and 
detectors placed at  ${\bf r}_1$ and ${\bf r}_2$, HBT interference 
produces fringes whose spatial variation is proportional to the 
expression $\cos^2[({\bf k}_1 - {\bf k}_2)({\bf r}_1-{\bf r}_2)/2]$ 
\cite{Scully1997}. If the maximum distance $|{\bf r}_1-{\bf r}_2|$, 
which can be viewed as the aperture of the measuring system, is fixed, 
an attempt to retrieve the angular separation between the sources from 
HBT fringes will suffer from the Rayleigh curse in the limit $|{\bf k}_1 
- {\bf k}_2| \rightarrow 0$. This is because for vanishing $|{\bf k}_1 - 
{\bf k}_2|$ one will observe only a small fraction of the HBT fringe in 
the vicinity of its maximum. 

In the case of the two-photon scheme presented here, we should emphasize the role of the 
prior QND measurement if superresolution is to be achieved with 
classical thermal light sources. While HBT interferometry works also 
with classical light sources, albeit with reduced visibility, the 
enhancement offered by our scheme stems from realizing two-photon 
interferometry sufficiently close to the dark fringe, i.e.\ with high 
visibility ${\cal V}$. In fact, since classical light sources can attain 
at most $50\%$ visibility of Hong-Ou-Mandel interference, formulas 
\eqref{eq:2p-1} indicate that no significant improvement is possible 
over the DI scheme:
for $\mathcal{V}=0.5$ we get
$(\Delta^2\varepsilon)^{-1}_{2\textrm{P}}= \frac{5\varepsilon^2}{32}$ 
vs. $(\Delta^2\varepsilon)^{-1}_{\textrm{DI}}= \frac{\varepsilon^2}{8}$ 
in case of direct imaging.

In conclusion, we have demonstrated both theoretically and experimentally an imaging protocol that circumvents the difficulties in a multi-parameter estimation problem by use of a collective measurement. The presented experimental results conclusively confirm the possibility
to exploit the inherent indistinguishability of photons to perform quantum-enhanced
simultaneous estimation of source separation and centroid. With this
proof-of-principle experiment we have also proposed a set of realistic
schemes in which our protocol could be readily applied, even to gain
additional information along the traditional single-photon DI scenario or other superresolution techniques.
The general theory of super-resolved imaging \cite{Rehacek2017, Chrostowski2017, Zhou2018} implies that the same protocol might be directly applied in case of a more general light source distribution provided one would be interested in estimating its first and second moments.

\begin{acknowledgements}
This work has been funded by the National Science Centre (Poland) Projects No. 2017/25/N/ST2/01163, 2016/22/E/ST2/00559 and by the project ``Quantum Optical Communication Systems'' carried out within the TEAM
programme of the Foundation for Polish Science co-financed by the European Union under the European Regional Development Fund. M.P. thanks M. Jachura for know-how transfer on the photon-pair source and M. Mazelanik for insightful discussions.
\end{acknowledgements}
\nocite{Chrapkiewicz2016}
\nocite{Lipka2018}
\bibliography{bibliografia}

\end{document}